\documentclass[final,5p,times,twocolumn]{elsarticle}

\usepackage{amsmath,amssymb,bm,mathtools}
\usepackage{slashed}
\usepackage{microtype}
\usepackage[colorlinks=true,allcolors=blue]{hyperref}

\allowdisplaybreaks[2]
\biboptions{sort&compress}

\newcommand{\dd}{\mathrm d}
\newcommand{\ii}{\mathrm i}
\newcommand{\lrD}{\overleftrightarrow D}
\newcommand{\ep}{\epsilon^{\prime *}}
\newcommand{\ME}[1]{\langle p',\lambda'|#1|p,\lambda\rangle}
\newcommand{\Kern}{\mathcal K}
\newcommand{\Ht}{\tilde H}
\newcommand{\epsT}{\epsilon_T}

\journal{Physics Letters B}

\begin{document}

\begin{frontmatter}

\title{Quark spin-orbit correlations in spin-1 targets}

\author[inha]{Hyunwoo Kim}
\author[inha,iqs]{June-Young Kim\corref{cor1}}
\cortext[cor1]{Corresponding author}
\ead{jun-young.kim@inha.ac.kr}
\address[inha]{Department of Physics, Inha University, Incheon 22212, Republic of Korea}
\address[iqs]{Institute of Quantum Science, Inha University, Incheon 22212, Republic of Korea}

\begin{abstract}
The quark spin-orbit correlation probes the alignment of quark helicity with longitudinal kinetic orbital angular momentum inside a hadron.  This correlation is defined by a QCD operator: the position moment of the asymmetric parity-odd quark energy-momentum tensor.  The matrix element of this rank-two tensor decomposes into symmetric-traceless, antisymmetric, and trace parts.  The symmetric-traceless part is matched to moments of axial generalized parton distributions.  The QCD equations of motion relate the antisymmetric part to vector and tensor form factors and set the trace to zero.  Using these relations, we derive two gauge-invariant sum rules for the spin-orbit correlation in a spin-1 hadron.  One gives the correlation in an unpolarized target.  The other gives its tensor-polarization dependence, which is absent for spin-0 and spin-$1/2$ targets.  We estimate the unpolarized spin-orbit correlations for the $\rho$ meson and deuteron using existing lattice and phenomenological inputs, respectively.
\end{abstract}
\end{frontmatter}
\section{Introduction}
\label{sec:intro}
A central question in hadron physics and a major focus of the Electron-Ion
Collider (EIC) program is how the intrinsic spin and orbital angular momentum
of quarks and gluons contribute to hadron spin~\cite{Accardi:2012qut}.
Correlations between these degrees of freedom are
equally essential to understanding the hadron's internal spin configuration.
The quark spin-orbit correlation probes this configuration by
quantifying the alignment of quark helicity with longitudinal kinetic
orbital angular momentum.

Kinetic quark orbital angular momentum is obtained from a moment of the
parity-even quark energy-momentum tensor (EMT).  When the quark field is
separated into right- and left-handed components, the moment of this tensor
adds the orbital angular momenta in the two chiral sectors, whereas the same
moment of its parity-odd counterpart takes their difference.  On the light
front, this chiral difference becomes the difference between the orbital
angular momenta carried by positive- and negative-helicity quarks.  It thus
weights orbital motion by quark spin and defines the quark spin-orbit
correlation.

This local, gauge-invariant correlation has previously been studied for
spin-0 and spin-$1/2$
targets~\cite{Lorce:2014mxa,Hatta:2024hqx,LorceSong:2025spin0,Kim:2024soc}.
The parity-odd EMT that defines it is an asymmetric rank-two tensor with
symmetric-traceless, antisymmetric, and trace parts.
The symmetric-traceless part is matched to moments of axial-vector
generalized parton distributions (GPDs)~\cite{Berger:2001zb,Cosyn:2018rdm},
while the QCD equations of motion express the antisymmetric part through
vector and tensor form factors and set the trace to zero.  The
correlation can therefore be expressed in terms of these GPD moments and form
factors.

We extend this analysis to spin-1 targets, whose polarization structure
includes tensor polarization, a quadrupole component present only for
targets with $J>1/2$.  In this work, we construct a complete covariant
parametrization of the parity-odd quark EMT and derive two gauge-invariant
sum rules that express, respectively, the unpolarized and tensor-polarized
quark spin-orbit correlations in terms of moments of the spin-1 axial-vector
GPDs and local vector and tensor form factors.  The unpolarized correlation
measures the alignment of quark helicity with orbital motion in a
spin-averaged target.  The tensor-polarized correlation characterizes the
response of this alignment to the target's quadrupole polarization.  We then
estimate the unpolarized correlations for the $\rho$ meson and deuteron using
existing lattice and phenomenological inputs, respectively.

\section{Parity-odd energy-momentum tensor and the spin-orbit correlation}
\label{sec:conventions}
We define the gauge-invariant, asymmetric, parity-odd quark EMT of flavor
$q$ as~\cite{Lorce:2014mxa}
\begin{align}
 \hat T_{q5}^{\mu\nu}(y)
 =\bar\psi_q(y)\gamma^\mu\gamma_5\ii\lrD^{\nu}\psi_q(y),
 \label{eq:T5def}
\end{align}
where $\psi_q(y)$ denotes the quark field of flavor $q$ at the spacetime
point $y$.  The covariant derivative entering this operator acts
bidirectionally and is defined by
\begin{align}
 \lrD^{\nu}
 \equiv\frac12\left(
 \overrightarrow D^{\nu}-\overleftarrow D^{\nu}
 \right).
\end{align}
We focus below on the quark operator.  The local gluonic counterpart is
discussed in Sec.~\ref{sec:traceEOM}.

The longitudinal quark spin-orbit correlation is then defined
at fixed light-front time $y^+=0$ by the
following position moment of the parity-odd EMT in
Eq.~\eqref{eq:T5def}~\cite{Lorce:2014mxa},
\begin{align}
 \hat C_z^q
 &=\int \dd y^-\dd^2\bm y_\perp\,
 \bar\psi_q(y)\gamma^+\gamma_5
 (\bm y\!\times\!\ii\lrD)_z\psi_q(y)
 \nonumber\\
 &=\int \dd y^-\dd^2\bm y_\perp\,
 \left[y^1\hat T_{q5}^{+2}(y)-y^2\hat T_{q5}^{+1}(y)\right],
 \label{eq:Coperator}
\end{align}
where we use $v^\mu=(v^+,v^-,\bm v_\perp)$, with light-front components
$v^\pm=(v^0\pm v^3)/\sqrt{2}$.

The expectation value of $\hat C_z^q$ in a hadron state measures the quark
spin-orbit correlation inside the target.  For a target of helicity
$\lambda$, we therefore define the correlation by
\begin{align}
 C_z^q(\lambda)
 =\frac{\langle P,\lambda|\hat C_z^q|P,\lambda\rangle}
 {\langle P,\lambda|P,\lambda\rangle}.
 \label{eq:Cdiag}
\end{align}
Here $|P,\lambda\rangle$ denotes a one-hadron state of mass $M$, momentum
$P$, and helicity $\lambda$, with $P^2=M^2$.  For general on-shell momenta
$k$ and $k'$, we use the light-front normalization
\begin{align}
 \langle k',\lambda'|k,\lambda\rangle
 =2k^+(2\pi)^3\delta(k^{\prime +}-k^+)
 \delta^{(2)}(\bm k'_{\perp}-\bm k_{\perp})
 \delta_{\lambda'\lambda}.
 \label{eq:statenormalization}
\end{align}
Importantly, $C_z^q(\lambda)$ is defined as a forward expectation
value, but the explicit position weight in $\hat C_z^q$ requires the nonforward
matrix element of $\hat T_{q5}^{\mu\nu}(0)$ near the forward limit.


\section{Covariant parametrization of the parity-odd EMT for a spin-1 target}
\label{sec:decomposition}
For a spin-1 target, the nonforward information required above is contained
in the parity-odd EMT matrix element between an initial state
$|p,\lambda\rangle$ and a final state $|p',\lambda'\rangle$.  With these
states normalized as in Eq.~\eqref{eq:statenormalization}, we write
\begin{align}
 \mathcal M_{q5}^{\mu\nu}(\lambda',\lambda;P,\Delta)
 &\equiv \ME{\hat T_{q5}^{\mu\nu}(0)}.
 \label{eq:nonforwardME}
\end{align}
The momentum dependence of $\mathcal M_{q5}^{\mu\nu}$ is expressed in
terms of the average hadron momentum $P$, the momentum transfer $\Delta$,
and the invariant momentum transfer $t$, defined by
\begin{equation}
 P=\frac{p'+p}{2},\qquad
 \Delta=p'-p,\qquad
 t=\Delta^2,
 \label{eq:kinematics}
\end{equation}
Both external momenta satisfy $p^2=p'^2=M^2$, where $M$ denotes the target
mass; these on-shell conditions imply $P\cdot\Delta=0$ and
$P^2=M^2-t/4$.  The remaining dependence on $\lambda$ and $\lambda'$ is
carried by the initial and final spin-1 polarization vectors, defined as
\begin{equation}
 \epsilon^\mu\equiv\epsilon^\mu(p,\lambda),\qquad
 \epsilon^{\prime *\mu}\equiv \epsilon^{*\mu}(p',\lambda').
 \label{eq:polvectors}
\end{equation}
These vectors satisfy $p\cdot\epsilon=p'\cdot\ep=0$, and their contraction
in the forward limit is
$\ep\cdot\epsilon=-\delta_{\lambda'\lambda}$.

The momentum variables in Eq.~\eqref{eq:kinematics} and the polarization
vectors in Eq.~\eqref{eq:polvectors} provide the building blocks for the
covariant parametrization of $\mathcal M_{q5}^{\mu\nu}$.  The corresponding
rank-four tensors carry the indices $\mu$, $\nu$, $\alpha'$, and $\alpha$
and are built from $P^\rho$, $\Delta^\rho$, $g^{\rho\sigma}$, and
$\epsilon^{\rho\sigma\kappa\eta}$.  The Lorentz indices $\alpha'$ and
$\alpha$ are contracted with the final- and initial-state polarization
vectors $\epsilon_{\alpha'}^{\prime *}$ and $\epsilon_\alpha$,
respectively.  Parity invariance, time-reversal
invariance, and hermiticity leave nine linearly independent covariant
tensors, which give the parametrization
\begin{align}
 \mathcal M_{q5}^{\mu\nu}(\lambda',\lambda;P,\Delta)
 &=\sum_{X=A,B,C,D}
 \Kern_X^{\mu\nu}(\lambda',\lambda;P,\Delta)F_X^q(t)
 \nonumber\\
 &\quad+\sum_{r=1}^{5}
 \Kern_{F_r}^{\mu\nu}(\lambda',\lambda;P,\Delta)F_r^q(t) .
 \label{eq:generalparam}
\end{align}
The covariant tensors $\Kern_X^{\mu\nu}$ and
$\Kern_{F_r}^{\mu\nu}$ carry the Lorentz and polarization structures,
whereas $F_X^q$ and $F_r^q$ are scalar form factors that depend on $t$.
The arguments displayed in Eq.~\eqref{eq:generalparam} are suppressed below.

We now organize the structures in Eq.~\eqref{eq:generalparam} into
irreducible Lorentz projections so that the constraints on each part can be
analyzed separately in Secs.~\ref{sec:gpdrelation}--\ref{sec:traceEOM}.
For this purpose, we use
$A^{\{\mu\nu\}}\equiv
\tfrac12(A^{\mu\nu}+A^{\nu\mu})-\tfrac14g^{\mu\nu}A_\alpha{}^\alpha$
and $A^{[\mu\nu]}\equiv\tfrac12(A^{\mu\nu}-A^{\nu\mu})$ for the
symmetric-traceless and antisymmetric projections, respectively.  The
parity-odd EMT then decomposes as
\begin{equation}
 \hat T_{q5}^{\mu\nu}(y)
 =\hat T_{q5}^{\{\mu\nu\}}(y)
 +\hat T_{q5}^{[\mu\nu]}(y)
 +\frac14g^{\mu\nu}g_{\alpha\beta}\hat T_{q5}^{\alpha\beta}(y).
 \label{eq:twistdecomp}
\end{equation}
This decomposition carries over directly to the matrix element in
Eq.~\eqref{eq:nonforwardME}, whose symmetric-traceless part contains four
covariant tensors
\begin{subequations}
\label{eq:covarianttensors}
\begin{align}
 \Kern_A^{\mu\nu}
 ={}&-2\ii\epsilon^{\{\mu\ep\epsilon P}P^{\nu\}},
 \\[0.5ex]
 \Kern_B^{\mu\nu}
 ={}&\frac{4\ii}{M^2}
 \left[
 \epsilon^{\{\mu\Delta P\epsilon}(\ep\!\cdot P)
 +\epsilon^{\{\mu\Delta P\ep}(\epsilon\!\cdot P)
 \right]P^{\nu\}},
 \\
 \Kern_C^{\mu\nu}
 ={}&\frac{4\ii}{M^2}
 \left[
 \epsilon^{\{\mu\Delta P\epsilon}(\ep\!\cdot P)
 -\epsilon^{\{\mu\Delta P\ep}(\epsilon\!\cdot P)
 \right]\Delta^{\nu\}},
 \\[0.5ex]
 \Kern_D^{\mu\nu}
 ={}&\ii\Big[
 \epsilon^{\{\mu\Delta P\epsilon}
 \epsilon^{\prime *\nu\}}
 +\epsilon^{\{\mu\Delta P\ep}
 \epsilon^{\nu\}}\Big],
\end{align}
\end{subequations}
whereas the antisymmetric part yields the remaining five covariant tensors
\begin{subequations}
\label{eq:covariantFtensors}
\begin{align}
 \Kern_{F_1}^{\mu\nu}
 &=-\frac{\ii M^2}{2\sqrt2}\epsilon^{\mu\nu\ep\epsilon},
 \\[0.5ex]
 \Kern_{F_2}^{\mu\nu}
 &=\ii\left[
 \epsilon^{\mu\nu\Delta\epsilon}(\ep\!\cdot P)
 +\epsilon^{\mu\nu\Delta\ep}(\epsilon\!\cdot P)
 \right],
 \\[0.5ex]
 \Kern_{F_3}^{\mu\nu}
 &=\ii\epsilon^{\mu\nu\Delta P}
 (\ep\!\cdot\epsilon),
 \\
 \Kern_{F_4}^{\mu\nu}
 &=\frac{\ii}{M^2}\epsilon^{\mu\nu\Delta P}
 (\ep\!\cdot P)(\epsilon\!\cdot P),
 \\[0.5ex]
 \Kern_{F_5}^{\mu\nu}
 &=-\ii\left[
 \epsilon^{\mu\nu P\epsilon}(\ep\!\cdot P)
 -\epsilon^{\mu\nu P\ep}(\epsilon\!\cdot P)
 \right].
\end{align}
\end{subequations}
Here and below, the Levi-Civita tensor follows the convention
$\epsilon_{0123}=+1$~\cite{Cosyn:2018rdm,Cosyn:2019aio}.  We use compact
contractions; for example, $\epsilon^{\mu\nu\Delta P}\equiv
\epsilon^{\mu\nu\alpha\beta}\Delta_\alpha P_\beta$.
The tensors in Eqs.~\eqref{eq:covarianttensors} and
\eqref{eq:covariantFtensors} correspond to geometric twists two and three,
respectively~\cite{Ji:2000id,Hagler:2004yt,Cosyn:2018rdm,Cotogno:2019vjb}.
The trace part corresponds to geometric twist four, and discrete
symmetries require it to vanish,
\begin{equation}
 \frac14g^{\mu\nu}g_{\alpha\beta}
 \ME{\hat T_{q5}^{\alpha\beta}(0)}=0.
 \label{eq:traceprojection}
\end{equation}
Accordingly, no independent trace form factor occurs.  A direct check from
the QCD equations of motion is given later in Sec.~\ref{sec:traceEOM}.


\section{Spin-orbit correlations from the parity-odd EMT form factors}

\label{sec:LFextraction}


We now extract the spin-orbit correlations defined in
Eq.~\eqref{eq:Cdiag} from the nonforward parity-odd EMT matrix element in
Eq.~\eqref{eq:generalparam}.  We perform this extraction in a symmetric
light-front frame, in which the initial and final targets have equal plus
momenta and opposite transverse momenta:
\begin{equation*}
 p^+=p'^+=P^+,
 \qquad
 \bm p_\perp=-\frac{\bm\Delta_\perp}{2},
 \qquad
 \bm p'_\perp=\frac{\bm\Delta_\perp}{2}.
\end{equation*}
Equivalently,
\begin{equation*}
 P_\perp=0,
 \qquad
 \Delta^+=0,
 \qquad
 t=-\bm\Delta_\perp^2.
\end{equation*}
The forward-state definition in Eq.~\eqref{eq:Cdiag} requires equal initial
and final helicities, $\lambda'=\lambda$.  We place the transverse origin at the center of the target and
identify $\bm y_\perp$ in Eq.~\eqref{eq:Coperator} with the impact parameter
$\bm b_\perp$.  The $y^-$ integration selects $\Delta^+=0$.  The remaining
two-dimensional Fourier transform gives an unambiguous transverse
distribution~\cite{Burkardt:2000za,Diehl:2002he}, which we define as
\begin{equation}
 \langle \hat T_{q5}^{+j}\rangle_{\lambda\lambda}(\bm b_\perp)
 \equiv\frac{1}{2P^+}
 \int\frac{\dd^2\bm\Delta_\perp}{(2\pi)^2}
 e^{-\ii\bm\Delta_\perp\cdot\bm b_\perp}
 \langle p',\lambda|\hat T_{q5}^{+j}(0)|p,\lambda\rangle,
 \label{eq:T5impact}
\end{equation}
where $\bm b_\perp$ is conjugate to $\bm\Delta_\perp$.  The spin-orbit
correlation can then be written as
\begin{align}
 C_z^q(\lambda)
 &=\int\dd^2\bm b_\perp\,
 \epsT^{ij}b_\perp^i
 \langle \hat T_{q5}^{+j}\rangle_{\lambda\lambda}(\bm b_\perp)
 \nonumber\\
 &=-\frac{\ii\epsT^{ij}}{2P^+}
 \left.\frac{\partial}{\partial\Delta_\perp^i}
 \langle p',\lambda|\hat T_{q5}^{+j}(0)|p,\lambda\rangle
 \right|_{\Delta=0}.
 \label{eq:projection}
\end{align}
The second line is obtained by integrating by parts with respect to
$\bm\Delta_\perp$; the factor $b_\perp^i$ then acts as
$-\ii\partial/\partial\Delta_\perp^i$ on the nonforward matrix element.
For the transverse contraction, we use $\epsT^{12}=+1$.  As discussed
below Eq.~\eqref{eq:statenormalization}, we indeed see that the forward
matrix element of $\hat C_z^q$ in Eq.~\eqref{eq:Cdiag} requires information
about the nonforward matrix element of $\hat T_{q5}^{\mu\nu}(0)$ near
$\Delta=0$.

We then compute the nonforward matrix element appearing in
Eq.~\eqref{eq:projection} from the parametrization in
Eq.~\eqref{eq:generalparam}, using the light-front polarization vectors
specified in \ref{app:projection}.  The resulting
$+j$ components of the covariant tensors in
Eqs.~\eqref{eq:covarianttensors} and \eqref{eq:covariantFtensors} are
collected in \ref{app:termwise-multipoles}.
The $\Delta_\perp$ derivative and the contraction with $\epsT^{ij}$ in
Eq.~\eqref{eq:projection} then yield two diagonal target-spin
structures,
\begin{equation}
 C_z^q(\lambda)
 =\left[
 C_{z,U}^q\mathbf1
 +\frac32C_{z,Q}^q\hat Q^{33}
 \right]_{\lambda\lambda}.
 \label{eq:Cmultipole}
\end{equation}
Here $\mathbf1$ gives the unpolarized contribution, whereas the longitudinal
quadrupole component $\hat Q^{33}$ encodes tensor polarization.  We thus
refer to $C_{z,U}^q$ and $C_{z,Q}^q$ as the unpolarized and
tensor-polarized spin-orbit correlations, respectively.  Both correlations
receive contributions from the symmetric-traceless and antisymmetric parts
of the parity-odd EMT.  In terms of the form factors in
Eq.~\eqref{eq:generalparam}, we find
\begin{subequations}
\label{eq:CCQff}
\begin{align}
 C_{z,U}^q
 &=\left[
 \frac23F_A^q
 -\frac{1}{6\sqrt2}F_1^q-F_3^q
 +\frac13F_5^q
 \right]_{t=0},
 \\
 C_{z,Q}^q
 &=\left[
 -\frac13\left(F_A^q+3F_D^q\right)
 +\frac{1}{3\sqrt2}F_1^q+\frac13F_5^q
 \right]_{t=0}.
\end{align}
\end{subequations}
From the parametrizations in Eqs.~\eqref{eq:covarianttensors} and
\eqref{eq:covariantFtensors}, we see that $F_A^q$ and $F_D^q$ in
Eq.~\eqref{eq:CCQff} come from the symmetric-traceless part, whereas
$F_1^q$, $F_3^q$, and $F_5^q$ come from the antisymmetric part.

To make explicit how $C_{z,U}^q$ and $C_{z,Q}^q$ combine for each target
helicity, we write $\hat Q^{33}$ in the helicity basis ordered as
$(+1,0,-1)$.  In this basis, it takes the diagonal
form~\cite{Varshalovich:1988ye}
\begin{equation}
 \left[\hat Q^{33}\right]_{\lambda'\lambda}
 =\frac13
 \begin{pmatrix}
  1&0&0\\
  0&-2&0\\
  0&0&1
 \end{pmatrix}_{\lambda'\lambda}.
 \label{eq:Q33weight}
\end{equation}
With this representation, Eq.~\eqref{eq:Cmultipole} gives
\begin{equation}
 C_z^q(\pm1)=C_{z,U}^q+\frac12C_{z,Q}^q,
 \qquad
 C_z^q(0)=C_{z,U}^q-C_{z,Q}^q,
 \label{eq:Chelicity}
\end{equation}
for each target helicity.
The normalized sum over these three target-helicity components cancels the
tensor-polarized contribution and leaves the unpolarized spin-orbit
correlation $C_{z,U}^q$, as expected,
\begin{equation}
 \frac13\sum_{\lambda=-1}^{1}C_z^q(\lambda)=C_{z,U}^q.
 \label{eq:Cunpolarizedaverage}
\end{equation}


\section{\texorpdfstring{Relations to axial
GPDs}{Relations to axial GPDs}}

\label{sec:gpdrelation}



We found in Eq.~\eqref{eq:CCQff} that both the symmetric-traceless and
antisymmetric parts of the parity-odd EMT contribute to the spin-orbit
correlations.  These contributions are related to moments of axial GPDs and
local form factors, respectively.  We begin with the symmetric-traceless
contribution.
The axial GPDs enter through the nonforward bilocal correlator along a
lightlike direction $n$, with $n^2=0$,
\begin{align}
 &\mathcal A^q_{\lambda'\lambda}(x,\xi,t) \cr
 &=\int\frac{\dd\kappa}{2\pi}
 e^{2\ii x(P\cdot n)\kappa}
 \ME{\bar\psi_q(-n\kappa)\slashed n\gamma_5
 \mathcal W\psi_q(n\kappa)},
 \label{eq:axialcorrelator}
\end{align}
where $\xi=-\Delta\cdot n/(2P\cdot n)$ denotes the skewness, $x$ represents
the average fraction of the hadron's light-cone momentum carried by the
active quark, and
$\mathcal W \equiv \mathcal W[-n\kappa,n\kappa]$ denotes the straight
Wilson line between the two quark fields.
For a spin-1 target, we parametrize this correlator in terms of four axial
GPDs following Refs.~\cite{Berger:2001zb,Cosyn:2018rdm}.  Suppressing the
common kinematic arguments $(x,\xi,t)$, we write
\begin{align}
 \mathcal A^q_{\lambda'\lambda}
 ={}&-\frac{\ii\epsilon^{n\ep\epsilon P}}{P\cdot n}\Ht_1^q
 \nonumber\\
 &+\frac{2\ii}{M^2P\cdot n}
 \left[
 \epsilon^{n\Delta P\epsilon}(\ep\!\cdot P)
 +\epsilon^{n\Delta P\ep}(\epsilon\!\cdot P)
 \right]\Ht_2^q
 \nonumber\\
 &+\frac{2\ii}{M^2P\cdot n}
 \left[
 \epsilon^{n\Delta P\epsilon}(\ep\!\cdot P)
 -\epsilon^{n\Delta P\ep}(\epsilon\!\cdot P)
 \right]\Ht_3^q
 \nonumber\\
 &+\frac{\ii}{2(P\cdot n)^2}
 \left[
 \epsilon^{n\Delta P\epsilon}(\ep\!\cdot n)
 +\epsilon^{n\Delta P\ep}(\epsilon\!\cdot n)
 \right]\Ht_4^q,
 \label{eq:axialGPDdecomp}
\end{align}
where a common renormalization scale $\mu$ is understood for all
scale-dependent quantities.  We suppress the scale argument below.


The $x$-weighted integral of the correlator in
Eq.~\eqref{eq:axialcorrelator} gives the matrix element of the local twist-2
axial-vector operator.  In our convention, this operator equals one half of
the symmetric-traceless part of the parity-odd
EMT~\cite{Cosyn:2018rdm}, so that
\begin{align}
 \int_{-1}^{1}\dd x\,x
 \mathcal A^q_{\lambda'\lambda}(x,\xi,t)
 &=\frac{n_\mu n_\nu}{2(P\cdot n)^2}
 \ME{\hat T_{q5}^{\{\mu\nu\}}(0)}.
 \label{eq:axialSecondMoment}
\end{align}
The same $x$-weighted integral applied to the parametrization in
Eq.~\eqref{eq:axialGPDdecomp} yields the standard relations between the
axial GPD moments and generalized form factors (GFFs)~\cite{Cosyn:2018rdm},
\begin{subequations}
\label{eq:twist2moments}
\begin{align}
 \int_{-1}^{1}\dd x\,x\Ht_1^q(x,\xi,t)
 &=\tilde A_{2,0}^q(t),
 \label{eq:twist2momentH1}
 \\
 \int_{-1}^{1}\dd x\,x\Ht_2^q(x,\xi,t)
 &=\tilde B_{2,0}^q(t),
 \\
 \int_{-1}^{1}\dd x\,x\Ht_3^q(x,\xi,t)
 &=-2\xi\,\tilde C_{2,1}^q(t),
 \\
 \int_{-1}^{1}\dd x\,x\Ht_4^q(x,\xi,t)
 &=\tilde D_{2,0}^q(t).
 \label{eq:twist2momentH4}
\end{align}
\end{subequations}
Through Eq.~\eqref{eq:axialSecondMoment}, we match the GFFs in
Eq.~\eqref{eq:twist2moments} to our parametrization in
Eqs.~\eqref{eq:generalparam} and \eqref{eq:covarianttensors} and obtain
\begin{equation}
 \left\{F_A^q,F_B^q,F_C^q,F_D^q\right\}(t)
 =\left\{\tilde A_{2,0}^q,\tilde B_{2,0}^q,
 \tilde C_{2,1}^q,\tilde D_{2,0}^q\right\}(t).
 \label{eq:symmetricGFFmatching}
\end{equation}
Among the four form factors matched in
Eq.~\eqref{eq:symmetricGFFmatching}, only $F_A^q$ and $F_D^q$ enter the
spin-orbit correlations in Eq.~\eqref{eq:CCQff}.  These form factors
correspond to the axial GPD moments in Eqs.~\eqref{eq:twist2momentH1} and
\eqref{eq:twist2momentH4}, respectively.


\section{Relations to local form factors}

\label{sec:formfactorrelations}



We now turn to the antisymmetric contribution to the spin-orbit correlations
in Eq.~\eqref{eq:CCQff}.  The following QCD relation connects the
antisymmetric parity-odd EMT to local vector and tensor
currents~\cite{Lorce:2014mxa},
\begin{align}
 \hat T_{q5}^{[\mu\nu]}(y)
 =\frac{m_q}{2}\hat{\mathcal O}_{qT5}^{\mu\nu}(y)
 -\frac14\epsilon^{\mu\nu\alpha\beta}
 \partial_\alpha\hat{\mathcal O}_{qV,\beta}(y),
 \label{eq:QCDidentity}
\end{align}
where these local vector and tensor currents are defined by
\begin{subequations}
\label{eq:localcurrents}
\begin{align}
 \hat{\mathcal O}_{qV}^\mu(y)
 &=\bar\psi_q(y)\gamma^\mu\psi_q(y),
 \label{eq:localvectorcurrent} \\[0.5ex]
 \hat{\mathcal O}_{qT5}^{\mu\nu}(y)
 &=\bar\psi_q(y)\ii\sigma^{\mu\nu}\gamma_5\psi_q(y).
 \label{eq:localtensorcurrent}
\end{align}
\end{subequations}
Here $\sigma^{\mu\nu}=\ii[\gamma^\mu,\gamma^\nu]/2$ and
$\gamma_5=\ii\gamma^0\gamma^1\gamma^2\gamma^3$.
The connection between the antisymmetric form factors $F_r^q$ in
Eq.~\eqref{eq:generalparam} and the vector and tensor form factors follows
by sandwiching Eq.~\eqref{eq:QCDidentity} between initial and final spin-1
states.  We first parametrize the matrix element of the vector current in
Eq.~\eqref{eq:localvectorcurrent} as~\cite{Arnold:1979cg,Cosyn:2018rdm},
\begin{align}
 \ME{\hat{\mathcal O}_{qV}^\mu(0)}
 ={}&-2(\ep\!\cdot\epsilon)P^\mu A_{1,0}^q(t)
 \nonumber\\
 &-\left[\epsilon^\mu(\ep\!\cdot\Delta)
 -\epsilon^{\prime *\mu}(\epsilon\!\cdot\Delta)\right]B_{1,0}^q(t)
 \nonumber\\
 &+\frac{P^\mu}{M^2}(\ep\!\cdot\Delta)
 (\epsilon\!\cdot\Delta)C_{1,0}^q(t).
 \label{eq:vectorcurrent}
\end{align}
We follow the notation of Ref.~\cite{Cosyn:2018rdm}, where
$A_{1,0}^q$, $B_{1,0}^q$, and $C_{1,0}^q$ denote the first moments of
the vector GPDs $H_1^q$, $H_2^q$, and $H_3^q$, respectively.
With the state normalization in Eq.~\eqref{eq:statenormalization},
$A_{1,0}^q(0)=N_q-N_{\bar q}$ gives the number of quarks of flavor $q$
minus the corresponding antiquarks in the target.  We next parametrize the spin-1 matrix element of the tensor current in
Eq.~\eqref{eq:localtensorcurrent} in terms of five form
factors~\cite{Cosyn:2018thq,Cosyn:2018rdm},
\begin{align}
 \ME{\hat{\mathcal O}_{qT5}^{\mu\nu}(0)}
 =\frac{2}{M} \Big[&
\Kern_{F_1}^{\mu\nu}A_{1,0}^{qT}(t)
+\Kern_{F_2}^{\mu\nu}D_{1,0}^{qT}(t)
 \nonumber\\
 +&\Kern_{F_3}^{\mu\nu}F_{1,0}^{qT}(t)+\Kern_{F_4}^{\mu\nu}G_{1,0}^{qT}(t)
 \nonumber\\
 +&\Kern_{F_5}^{\mu\nu}I_{1,0}^{qT}(t)\Big].
 \label{eq:tensorcurrent}
\end{align}
This matrix element has the same five covariant tensors as that of
$\hat T_{q5}^{[\mu\nu]}$.  This is clear from the QCD relation in
Eq.~\eqref{eq:QCDidentity}.

We now compare the matrix elements on both sides of
Eq.~\eqref{eq:QCDidentity}.  Substituting the current parametrizations in
Eqs.~\eqref{eq:vectorcurrent} and \eqref{eq:tensorcurrent} into its
right-hand side and comparing the result with the antisymmetric parity-odd
EMT parametrization in Eq.~\eqref{eq:generalparam}, we obtain
\begin{subequations}
\label{eq:EOMrelations}
\begin{align}
 F_1^q(t)
 &=\frac{m_q}{M}A_{1,0}^{qT}(t),
 \\
 F_2^q(t)
 &=-\frac12B_{1,0}^q(t)
 +\frac{m_q}{M}D_{1,0}^{qT}(t),
 \\
 F_3^q(t)
 &=\frac12A_{1,0}^q(t)
 +\frac{m_q}{M}F_{1,0}^{qT}(t),
 \\
 F_4^q(t)
 &=C_{1,0}^q(t)
 +\frac{m_q}{M}G_{1,0}^{qT}(t),
 \\
 F_5^q(t)
 &=\frac{m_q}{M}I_{1,0}^{qT}(t).
\end{align}
\end{subequations}
The tensor-current contributions in
Eq.~\eqref{eq:EOMrelations} are all proportional to $m_q/M$, as follows from
the first term of Eq.~\eqref{eq:QCDidentity}. The vector current, in contrast, enters
Eq.~\eqref{eq:QCDidentity} through a total derivative.  Its contribution to
Eq.~\eqref{eq:EOMrelations} is obtained by applying this derivative to the
nonforward vector-current matrix element.  To apply it, we express the
current at $y$ through the four-momentum operator $\hat P^\mu$, which
generates spacetime translations,
\begin{align}
 \hat{\mathcal O}_{qV,\beta}(y)
 =e^{\ii\hat P\cdot y}\hat{\mathcal O}_{qV,\beta}(0)
 e^{-\ii\hat P\cdot y}.
\end{align}
Because the spin-1 states $|p,\lambda\rangle$ and $|p',\lambda'\rangle$ are
eigenstates of $\hat P^\mu$ with eigenvalues $p^\mu$ and $p'^\mu$,
respectively, differentiation of the vector-current matrix element with
respect to $y^\alpha$ at $y=0$ gives
\begin{align}
 \left.\frac{\partial}{\partial y^\alpha}
 \ME{\hat{\mathcal O}_{qV,\beta}(y)}\right|_{y=0}
 =\ii\Delta_\alpha\ME{\hat{\mathcal O}_{qV,\beta}(0)}.
\label{eq:vectorcurrentderivative}
\end{align}
Combining this derivative relation with the parametrization in
Eq.~\eqref{eq:vectorcurrent}, we obtain the vector-current contribution to
Eq.~\eqref{eq:EOMrelations}.  This derivative structure also determines
which vector form factor survives in the spin-orbit correlations.  The
$\Delta$ derivative in Eq.~\eqref{eq:projection} gives a nonzero term only
when it acts on the explicit $\Delta_\alpha$ in
Eq.~\eqref{eq:vectorcurrentderivative}.  If it acts instead on the $\Delta$
dependence of the vector-current matrix element, that factor remains and
forces the term to vanish at $\Delta=0$.  Only the forward vector-current
matrix element therefore survives.  The $B_{1,0}^q$ and $C_{1,0}^q$
structures in Eq.~\eqref{eq:vectorcurrent} vanish in this limit, leaving
\begin{align}
 \left.\ME{\hat{\mathcal O}_{qV}^{\mu}(0)}\right|_{\Delta=0}
 =2P^\mu A_{1,0}^q(0)\delta_{\lambda'\lambda}.
 \label{eq:forwardvectorcurrent}
\end{align}
It is proportional to the target-spin identity and thus contributes to
$C_{z,U}^q$, but not to $C_{z,Q}^q$.


\section{Trace of the parity-odd EMT}
\label{sec:traceEOM}


The preceding two sections related the symmetric-traceless and
antisymmetric parts of the parity-odd EMT to GPD moments and local form
factors, respectively.  The trace is the only remaining part.  Discrete
symmetries already required it to vanish in Eq.~\eqref{eq:traceprojection},
leaving no independent trace form factor.  We now close the decomposition
by confirming this result directly from the quark equations of motion.  With
$\{\gamma_5,\gamma^\mu\}=0$,
$\ii\slashed D\psi_q(y)=m_q\psi_q(y)$, and
$\bar\psi_q(y)\ii\overleftarrow{\slashed D}=-m_q\bar\psi_q(y)$, the operator
trace becomes
\begin{align}
 g_{\mu\nu}\hat T_{q5}^{\mu\nu}(y)
 &=\frac12\left[
 \bar\psi_q(y)\gamma^\mu\gamma_5\ii\overrightarrow D_\mu\psi_q(y)
 -\bar\psi_q(y)\ii\overleftarrow D_\mu\gamma^\mu\gamma_5\psi_q(y)
 \right]
 \nonumber\\
 &=\frac12(-m_q+m_q)\bar\psi_q(y)\gamma_5\psi_q(y)=0.
 \label{eq:tracezero}
\end{align}
Because the same mass $m_q$ appears in both terms of
Eq.~\eqref{eq:tracezero}, it cancels whether $q$ is light or heavy.  This
cancellation requires neither the chiral limit nor exact flavor symmetry
and leaves no independent trace form factor.  The result agrees with
Eq.~\eqref{eq:traceprojection}, obtained independently from the covariant
parametrization.

This conclusion applies to the flavor-diagonal matrix element parametrized
in Eq.~\eqref{eq:generalparam}.  A flavor-changing operator instead connects
different hadrons and is described by a separate transition parametrization.
Hermiticity then relates this matrix element to its conjugate in the reverse
process rather than constraining the same matrix element as in the
flavor-diagonal case.  At the operator level, the equations of motion give
$\tfrac12(m_{q'}-m_q)\bar\psi_{q'}(y)\gamma_5\psi_q(y)$, which vanishes only in
the equal-mass limit.  When $m_{q'}\ne m_q$, the pseudoscalar transition
operator governs the trace, and a corresponding trace form factor can
survive.  Such flavor-changing transitions lie outside the scope of the
present work.

The trace analysis also determines whether a local gluon operator can
generate an analogous spin-orbit correlation.  The possible gluonic
counterpart is a gauge-invariant rank-two operator constructed from the field
strength $F^{\mu\nu}$ and its dual $\widetilde F^{\mu\nu}$.  In four
dimensions, the Schouten identity fixes its Lorentz structure
as~\cite{LorceSong:2025spin0}
\begin{align}
 \hat T_{g5}^{\mu\nu}(y)
 &\equiv \ii\widetilde F^{\mu\alpha}(y)F_{\alpha}{}^\nu(y)
 \nonumber\\
 &=-\frac{\ii}{4}g^{\mu\nu}
 \widetilde F^{\alpha\beta}(y)F_{\alpha\beta}(y),
 \label{eq:gluonSchouten}
\end{align}
where $\widetilde F^{\mu\nu}
=\tfrac12\epsilon^{\mu\nu\alpha\beta}F_{\alpha\beta}$ and the color
contraction is understood. This operator obeys the same Lorentz and
discrete-symmetry constraints as the quark parity-odd EMT, so its spin-1
matrix element is restricted to the covariant structures in
Eq.~\eqref{eq:generalparam}.  The Schouten identity in
Eq.~\eqref{eq:gluonSchouten} makes the operator purely trace, whereas the
trace projection of the general parametrization vanishes in
Eq.~\eqref{eq:traceprojection}.  Its matrix element therefore vanishes,
\begin{align}
 \ME{\hat T_{g5}^{\mu\nu}(0)}=0.
 \label{eq:gluonMEzero}
\end{align}
Thus, the local parity-odd gluon EMT does not contribute to the spin-orbit
correlation in a spin-1 hadron.  The same conclusion was obtained for a
spin-0 target in Ref.~\cite{LorceSong:2025spin0}.


\section{Sum rules for spin-orbit correlations and numerical estimates}

\label{sec:sumrules}



\subsection{Sum rules for unpolarized and tensor-polarized targets}


The light-front analysis led to Eq.~\eqref{eq:CCQff}, where the unpolarized
and tensor-polarized spin-orbit correlations are expressed in terms of the
parity-odd EMT form factors.
The symmetric-traceless form factors are fixed by axial GPD moments in
Eq.~\eqref{eq:twist2moments}, and the antisymmetric ones by local vector and
tensor form factors in Eq.~\eqref{eq:EOMrelations}.  The trace analysis in
Sec.~\ref{sec:traceEOM} leaves no additional form factor.  We now combine
these relations with Eq.~\eqref{eq:CCQff} to obtain the two central sum
rules:
\begin{subequations}
\label{eq:master}
\begin{align}
 C_{z,U}^q
 ={}&\frac23\int_{-1}^{1}\dd x\,x
 \Ht_1^q(x,0,0)
 -\frac12A_{1,0}^q(0)
 \nonumber\\
 &+\frac{m_q}{M}\Bigg[
 -\frac{1}{6\sqrt2}A_{1,0}^{qT}(0)
 -F_{1,0}^{qT}(0)
 +\frac13I_{1,0}^{qT}(0)
 \Bigg],
 \label{eq:Cmaster}
 \\
 C_{z,Q}^q
 ={}&-\frac13\int_{-1}^{1}\dd x\,x
 \left[
 \Ht_1^q(x,0,0)+3\Ht_4^q(x,0,0)
 \right]
 \nonumber\\
 &+\frac{m_q}{3M}\left[
 \frac{1}{\sqrt2}A_{1,0}^{qT}(0)
 +I_{1,0}^{qT}(0)
 \right].
 \label{eq:CQmaster}
\end{align}
\end{subequations}
The renormalization-scale argument $\mu$, common to all scale-dependent
quantities in Eq.~\eqref{eq:master}, is suppressed.
The unpolarized spin-orbit correlation $C_{z,U}^q$ is expressed as the sum
of the $x$-weighted moment of $\Ht_1^q$, the vector charge
$A_{1,0}^q(0)$, and tensor-current contributions.  Here
$\Ht_1^q(x,0,0)$ reduces to the
quark helicity distribution 
$\Delta q(x)$ for $x>0$, while
$x<0$ accounts for antiquarks~\cite{Berger:2001zb}.  By contrast, the tensor-polarized
spin-orbit correlation $C_{z,Q}^q$ is expressed as the sum of the
$x$-weighted moment of $\Ht_1^q+3\Ht_4^q$ and tensor-current contributions.
In this case, the vector current does not contribute because the forward
vector-current matrix element in Eq.~\eqref{eq:forwardvectorcurrent} is
proportional to the target-spin identity. 

The tensor-current form factors remain in both correlations in
Eq.~\eqref{eq:master}.  According to the QCD
relation~\eqref{eq:QCDidentity}, their contributions always carry a factor
$m_q/M$.  Since $m_q/M\ll1$ for the $u$ and $d$ quarks, these terms can be
neglected; the same
approximation is not generally justified for heavy quarks.  With the
tensor-current terms omitted, the sum rules reduce to
\begin{subequations}
\label{eq:chirallimit}
\begin{align}
 C_{z,U}^q
 &=\frac23\int_{-1}^{1}\dd x\,x
 \Ht_1^q(x,0,0)
 -\frac12A_{1,0}^q(0),
 \\
 C_{z,Q}^q
 &=-\frac13\int_{-1}^{1}\dd x\,x
 \left[\Ht_1^q(x,0,0)+3\Ht_4^q(x,0,0)\right].
\end{align}
\end{subequations}
The reduced expressions in Eq.~\eqref{eq:chirallimit} can also be understood
as the chiral limit $m_q\to0$ of the general sum rules in
Eq.~\eqref{eq:master}.  Accordingly, we use them below to estimate the
$u$- and $d$-quark contributions to the $\rho$ meson and deuteron.

\subsection{\texorpdfstring{Numerical
estimates for the $\rho$ meson}{Numerical estimates for the rho meson}}
An earlier quenched lattice calculation determined the reduced matrix
element of the local twist-2 axial-vector operator for the $\rho$
meson~\cite{Best:1997qp}.  Through Eq.~\eqref{eq:axialSecondMoment}, this
result gives the forward-limit $x$-weighted moment of the axial GPD
$\Ht_1^u$.  That moment is needed in Eq.~\eqref{eq:chirallimit} to estimate
the unpolarized $u$-quark spin-orbit correlation in the $\rho^+$ meson,
$C_{z,U}^{u,\rho^+}$.
The chirally extrapolated value is quoted at $\mu=2.4~\mathrm{GeV}$ in the
$\overline{\mathrm{MS}}$ scheme.  Since the quenched calculation contains no
$\bar u$ sea contribution, its single-flavor result can be identified with
the valence $u$-quark moment in our convention,
\begin{equation}
 \int_{-1}^{1}\dd x\,x\Ht_1^{u,\rho^+}(x,0,0)
 =0.212(17).
 \label{eq:rhoaxialmoment}
\end{equation}
The $\rho^+$ carries unit net $u$-flavor charge, which fixes
$A_{1,0}^u(0)=1$.  We insert this charge and the lattice
moment in Eq.~\eqref{eq:rhoaxialmoment} into Eq.~\eqref{eq:chirallimit} and
obtain
\begin{equation}
 C_{z,U}^{u,\rho^+}\simeq-0.36.
 \label{eq:rhoestimate}
\end{equation}
After leading-order evolution to the same scale, the light-front calculation of
Ref.~\cite{Sun:2018bda} gives approximately $0.29$ for the same axial GPD
moment, which yields $C_{z,U}^{u,\rho^+}\simeq-0.31$.  Both estimates favor
antialignment between the valence $u$-quark helicity and its kinetic orbital
angular momentum.  The forward-limit $x$-weighted moment of
$\Ht_4^{u,\rho^+}$ has not been reported in
existing calculations~\cite{Sun:2018bda,Zhang:2022odc}, so
$C_{z,Q}^{u,\rho^+}$ cannot yet be estimated.

\subsection{\texorpdfstring{Numerical estimate for the deuteron}{Numerical estimate for the deuteron}}
The impulse-approximation calculation of the deuteron
axial GPDs suggests a small forward-limit $x$-weighted moment of
$\Ht_1^q$~\cite{Cano:2003ju}.  We therefore neglect this contribution in
Eq.~\eqref{eq:chirallimit}, leaving the vector charge as the dominant term.
For the deuteron $D$, the net $u$- and $d$-flavor charges are both
three, $A_{1,0}^{u,D}(0)=A_{1,0}^{d,D}(0)=3$.  We then obtain
\begin{equation}
 C_{z,U}^{u,D}=C_{z,U}^{d,D}
 \simeq -\frac32.
\label{eq:deuteronestimate}
\end{equation}
The equality of the two flavor contributions makes the nonsinglet
combination vanish.  The resulting singlet dominance is consistent with the
large-$N_c$ hierarchy found for the nucleon, where the flavor-singlet
spin-orbit correlation is leading and its nonsinglet counterpart is
suppressed~\cite{Kim:2024soc}.  For each light-quark flavor, the magnitude of
$C_{z,U}^{q,D}$ is twice that of the corresponding isoscalar-nucleon
correlation, as expected when the proton and neutron contributions add in
impulse approximation.  Its negative sign indicates a
preference for antialignment between light-quark helicity and kinetic orbital
angular momentum in the unpolarized deuteron.  By contrast, the
tensor-polarized correlation $C_{z,Q}^{q,D}$ cannot yet be estimated because
the forward-limit $x$-weighted moment of $\Ht_4^{q,D}$ is not available.


\section{Summary and outlook}
\label{sec:conclusion}
We have studied the kinetic quark spin-orbit correlation defined by the
forward expectation value of a light-front position moment of the asymmetric
parity-odd quark energy-momentum tensor (EMT).  This operator weights the
quark kinetic orbital motion by its helicity and thereby measures their
alignment.  To evaluate this correlation for a spin-1 hadron, we constructed the most
general nonforward matrix element of the parity-odd EMT.  Its nine independent
form factors divide into symmetric-traceless and antisymmetric parts, while no
independent trace form factor is allowed
for a flavor-diagonal matrix element.  The absence of a trace form factor
agrees with the QCD equations of motion and also excludes a
contribution from the local gluon counterpart, which the Schouten identity
makes purely trace.
Applying the position moment to this parametrization gives two target-spin
combinations.  The spin-independent term gives the unpolarized correlation
$C_{z,U}^q$, whereas the longitudinal quadrupole term gives the
tensor-polarized correlation $C_{z,Q}^q$. The tensor-polarized correlation has no counterpart for
spin-0 or spin-$1/2$ targets.

We next related these correlations to axial generalized parton distributions
(GPDs) and local currents.  The symmetric-traceless part matches second
moments of the axial GPDs, while an exact QCD relation connects the
antisymmetric part to vector and tensor form factors.  Together, these
relations give a separate gauge-invariant sum rule for each correlation.
The sum rules show that
$C_{z,U}^q$ depends on the $x$-weighted moment of $\Ht_1^q$ and the vector
charge.  By contrast, $C_{z,Q}^q$ probes the moment of
$\Ht_1^q+3\Ht_4^q$, with no contribution from the vector current.  The tensor
terms in both relations are proportional to $m_q/M$ and are negligible for
the light quarks.

Existing lattice and light-front inputs give
$C_{z,U}^{u,\rho^+}\simeq-0.36$ and $-0.31$, respectively, for the valence
$u$ quark in the $\rho^+$.  Both estimates are negative and favor
antialignment between its helicity and kinetic orbital angular momentum.
For the deuteron, the impulse-approximation estimate yields
$C_{z,U}^{u,D}=C_{z,U}^{d,D}\simeq-3/2$.  The common negative sign favors
antialignment for both light-quark flavors.  The tensor-polarized correlation
remains unknown because the forward-limit $x$-weighted moment of $\Ht_4^q$
is not available for either system.  Determining this moment, or more
generally the $x$ dependence of $\Ht_4^q$, is therefore the most direct next
step.  Its dynamical interpretation also remains to be clarified.

Several directions remain open.  The finite-$t$ dependence of the form
factors would map the
transverse spatial distributions of the unpolarized and tensor-polarized
correlations.  Heavy flavors pose a different problem because the tensor
terms proportional to $m_q/M$ need not be small.  Flavor-changing
transitions also require their own parametrization.  In that case, the
symmetric-traceless form factors are moments of transition GPDs, while
unequal quark masses allow a pseudoscalar trace contribution.

\enlargethispage{2\baselineskip}
\section*{Acknowledgments}
H.K. thanks Do Wan Kim and Ulugbek Yakhshiev for their support.
J.-Y.K. thanks C\'{e}dric Lorc\'{e} for invaluable discussions and
the Centre de Physique Th\'{e}orique (CPHT) at \'{E}cole Polytechnique
for its hospitality during his visit. 
He also thanks Hyun-Chul Kim for his longstanding support and encouragement.

\appendix

\section{Light-front polarizations and target-spin multipoles}
\label{app:projection}
In the target-spin space, we choose the three-dimensional polarization
vectors
\begin{align}
 \bm\varepsilon(\pm)=\mp\frac{1}{\sqrt2}(1,\pm \ii,0), \qquad 
 \bm\varepsilon(0)=(0,0,1).
 \label{eq:spinpolarizations}
\end{align}
Here $\bm\varepsilon_\perp$ denotes the first two components of
$\bm\varepsilon$, and
$\varepsilon_a^*(\lambda')\varepsilon_a(\lambda)
=\delta_{\lambda'\lambda}$ for $a=1,2,3$.  In the symmetric light-front
frame of Sec.~\ref{sec:LFextraction}, the corresponding helicity vectors
are
\begin{subequations}
\label{eq:LFpol}
\begin{align}
 \epsilon^\mu(p,\pm)
 &=\left(
 0,\frac{\bm p_\perp\!\cdot\bm\varepsilon_\perp(\pm)}{p^+},
 \bm\varepsilon_\perp(\pm)\right),
 \\
 \epsilon^\mu(p,0)
 &=\left(
 \frac{p^+}{M},
 \frac{\bm p_\perp^2-M^2}{2Mp^+},
 \frac{\bm p_\perp}{M}\right).
\end{align}
\end{subequations}
The corresponding vectors for the final state are obtained by $p\to p'$
and complex conjugation.

Using the three-dimensional polarization vectors in
Eq.~\eqref{eq:spinpolarizations}, we define the helicity-basis matrix
elements of the spin-1 multipole operators as
\begin{subequations}
\label{eq:SQpolarization}
\begin{align}
 (\hat S^a)_{\lambda'\lambda}
 &=-\ii\epsilon^{abc}
 \varepsilon_b^*(\lambda')\varepsilon_c(\lambda),
 \label{eq:SQpolarization-a}
 \\[0.5ex]
 (\hat Q^{ab})_{\lambda'\lambda}
 &=\frac13\delta^{ab}\delta_{\lambda'\lambda}-\frac12\left[
 \varepsilon_a^*(\lambda')\varepsilon_b(\lambda)
 +\varepsilon_b^*(\lambda')\varepsilon_a(\lambda)\right].
 \label{eq:SQpolarization-b}
\end{align}
\end{subequations}

\section{Target-spin decomposition of the covariant tensors}
\label{app:termwise-multipoles}
Suppressing the common helicity indices $\lambda'\lambda$, the $+i$
components of the covariant tensors in Eq.~\eqref{eq:generalparam},
divided by $2P^+$, are listed as follows:
\begin{subequations}
\label{eq:fullspinCoefficients}
\begin{align}
 \frac{\Kern_A^{+i}}{2P^+}
 ={}&-\frac{\ii}{3}\epsT^{ij}\Delta_\perp^j
 \mathbf1
 +\left(\frac{M}{2}+\frac{t}{8M}\right)
 \hat S^i
 \nonumber\\
 &+\frac{\Delta_\perp^i\Delta_\perp^j}{8M}
 \hat S^j
 -\frac{\ii}{2}\epsT^{jk}\Delta_\perp^k
 \hat Q^{ij},
 \label{eq:fullspinKA}
 \\
 \frac{\Kern_B^{+i}}{2P^+}
 ={}&+\frac{\ii t}{3M^2}
 \epsT^{ij}\Delta_\perp^j\mathbf1-\frac{4M^2+t}{8M^3}
 \left[t\hat S^i
 +\Delta_\perp^i\Delta_\perp^j
 \hat S^j\right]
 \nonumber\\
 &+\frac{\ii}{2M^2}\epsT^{ij}\Delta_\perp^j
 \left[\Delta_\perp^k\Delta_\perp^l
 \hat Q^{kl}-t\hat Q^{33}\right],
 \label{eq:fullspinKB}
 \\
 \frac{\Kern_C^{+i}}{2P^+}
 ={}&+\frac{t}{4M^3}
 \Delta_\perp^i\Delta_\perp^j
 \hat S^j
-\frac{\ii}{M^2}
 \Delta_\perp^i\Delta_\perp^j
 \epsT^{kl}\Delta_\perp^l
 \hat Q^{jk},
 \label{eq:fullspinKC}
 \\
 \frac{\Kern_D^{+i}}{2P^+}
 ={}&+\frac{t}{8M}\hat S^i
 -\frac{\ii}{2}\epsT^{jk}\Delta_\perp^k
 \hat Q^{ij}+\frac{\ii}{2}\epsT^{ij}\Delta_\perp^j
 \hat Q^{33}.
 \label{eq:fullspinKD}
\end{align}
\begin{align}
 \frac{\Kern_{F_1}^{+i}}{2P^+}
 ={}&+\frac{\ii}{12\sqrt2}\epsT^{ij}\Delta_\perp^j
 \mathbf1
 -\frac{M}{4\sqrt2}\hat S^i
 -\frac{\ii}{4\sqrt2}\epsT^{ij}\Delta_\perp^j
 \hat Q^{33},
 \label{eq:fullspinKF1}
 \\
 \frac{\Kern_{F_2}^{+i}}{2P^+}
 ={}&+\frac{\ii t}{12M^2}
 \epsT^{ij}\Delta_\perp^j\mathbf1-\frac{1}{4M}
 \left[t\hat S^i
 +\Delta_\perp^i\Delta_\perp^j
 \hat S^j\right]
 \nonumber\\
 &-\frac{\ii t}{4M^2}
 \epsT^{ij}\Delta_\perp^j
 \hat Q^{33},
 \label{eq:fullspinKF2}
 \\
 \frac{\Kern_{F_3}^{+i}}{2P^+}
 ={}&+\frac{\ii(6M^2+t)}{12M^2}
 \epsT^{ij}\Delta_\perp^j\mathbf1
-\frac{1}{2M}
 \left[t\hat S^i
 +\Delta_\perp^i\Delta_\perp^j
 \hat S^j\right]
 \nonumber\\
 &-\frac{\ii t}{4M^2}
 \epsT^{ij}\Delta_\perp^j
 \hat Q^{33},
 \label{eq:fullspinKF3}
 \\
 \frac{\Kern_{F_4}^{+i}}{2P^+}
 ={}&-\frac{\ii t(4M^2+t)}{96M^4}
 \epsT^{ij}\Delta_\perp^j\mathbf1
+\frac{t}{16M^3}
 \left[t\hat S^i
 +\Delta_\perp^i\Delta_\perp^j
 \hat S^j\right]
 \nonumber\\
 &-\frac{\ii}{8M^2}
 \epsT^{ij}\Delta_\perp^j
 \Delta_\perp^k\Delta_\perp^l
 \hat Q^{kl}+\frac{\ii t^2}{32M^4}
 \epsT^{ij}\Delta_\perp^j
 \hat Q^{33},
 \label{eq:fullspinKF4}
 \\
 \frac{\Kern_{F_5}^{+i}}{2P^+}
 ={}&-\frac{\ii(4M^2+t)}{24M^2}
 \epsT^{ij}\Delta_\perp^j\mathbf1
 +\frac{t}{4M}\hat S^i
+\frac{\Delta_\perp^i\Delta_\perp^j}{8M}
 \hat S^j
  \nonumber\\
 &-\frac{\ii}{2}\epsT^{jk}\Delta_\perp^k
 \hat Q^{ij}-\frac{\ii(4M^2-t)}{8M^2}
 \epsT^{ij}\Delta_\perp^j
 \hat Q^{33}.
 \label{eq:fullspinKF5}
\end{align}
\end{subequations}

\bibliographystyle{elsarticle-num}
\bibliography{spin1_spin_orbit_plb_final}

@article{Lorce:2014mxa,
  author        = {Lorc{\'e}, C{\'e}dric},
  title         = {Spin--orbit correlations in the nucleon},
  journal       = {Phys. Lett. B},
  volume        = {735},
  pages         = {344--348},
  year          = {2014},
  eprint        = {1401.7784},
  archiveprefix = {arXiv},
  primaryclass  = {hep-ph}
}

@article{LorceSong:2025spin0,
  author        = {Lorc{\'e}, C{\'e}dric and Song, Qin-Tao},
  title         = {Spin-orbit correlation and spatial distributions for spin-0 hadrons},
  journal       = {Phys. Lett. B},
  volume        = {864},
  pages         = {139433},
  year          = {2025},
  doi           = {10.1016/j.physletb.2025.139433},
  eprint        = {2501.05092},
  archiveprefix = {arXiv},
  primaryclass  = {hep-ph}
}

@article{Accardi:2012qut,
  author        = {Accardi, A. and others},
  title         = {Electron Ion Collider: The Next {QCD} Frontier---Understanding the glue that binds us all},
  journal       = {Eur. Phys. J. A},
  volume        = {52},
  number        = {9},
  pages         = {268},
  year          = {2016},
  doi           = {10.1140/epja/i2016-16268-9},
  eprint        = {1212.1701},
  archiveprefix = {arXiv},
  primaryclass  = {nucl-ex}
}

@article{Hatta:2024hqx,
  author        = {Hatta, Yoshitaka and Schoenleber, Sebastian},
  title         = {Twist analysis of the spin-orbit correlation in {QCD}},
  journal       = {J. High Energy Phys.},
  volume        = {09},
  pages         = {154},
  year          = {2024},
  eprint        = {2404.18872},
  archiveprefix = {arXiv},
  primaryclass  = {hep-ph}
}

@article{Berger:2001zb,
  author        = {Berger, E. R. and Cano, F. and Diehl, M. and Pire, B.},
  title         = {Generalized parton distributions in the deuteron},
  journal       = {Phys. Rev. Lett.},
  volume        = {87},
  pages         = {142302},
  year          = {2001},
  eprint        = {hep-ph/0106192},
  archiveprefix = {arXiv}
}

@article{Cano:2003ju,
  author        = {Cano, F. and Pire, B.},
  title         = {Deep electroproduction of photons and mesons on the deuteron},
  journal       = {Eur. Phys. J. A},
  volume        = {19},
  pages         = {423--438},
  year          = {2004},
  doi           = {10.1140/epja/i2003-10127-x},
  eprint        = {hep-ph/0307231},
  archiveprefix = {arXiv}
}

@article{Cosyn:2018thq,
  author        = {Cosyn, Wim and Pire, Bernard},
  title         = {Transversity generalized parton distributions for the deuteron},
  journal       = {Phys. Rev. D},
  volume        = {98},
  pages         = {074020},
  year          = {2018},
  eprint        = {1806.01177},
  archiveprefix = {arXiv},
  primaryclass  = {hep-ph}
}

@article{Cosyn:2018rdm,
  author        = {Cosyn, Wim and Freese, Adam and Pire, Bernard},
  title         = {Polynomiality sum rules for generalized parton distributions of spin-1 targets},
  journal       = {Phys. Rev. D},
  volume        = {99},
  pages         = {094035},
  year          = {2019},
  eprint        = {1812.01511},
  archiveprefix = {arXiv},
  primaryclass  = {hep-ph}
}

@article{Cosyn:2019aio,
  author        = {Cosyn, Wim and Cotogno, Sabrina and Freese, Adam and Lorc{\'e}, C{\'e}dric},
  title         = {The energy--momentum tensor of spin-1 hadrons: Formalism},
  journal       = {Eur. Phys. J. C},
  volume        = {79},
  pages         = {476},
  year          = {2019},
  eprint        = {1903.00408},
  archiveprefix = {arXiv},
  primaryclass  = {hep-ph}
}

@article{Cotogno:2019vjb,
  author        = {Cotogno, Sabrina and Lorc{\'e}, C{\'e}dric and Lowdon, Peter and Morales, Manuel},
  title         = {Covariant multipole expansion of local currents for massive states of any spin},
  journal       = {Phys. Rev. D},
  volume        = {101},
  pages         = {056016},
  year          = {2020},
  eprint        = {1912.08749},
  archiveprefix = {arXiv},
  primaryclass  = {hep-ph}
}

@article{Arnold:1979cg,
  author  = {Arnold, R. G. and Carlson, C. E. and Gross, F.},
  title   = {Elastic electron--deuteron scattering at high energy},
  journal = {Phys. Rev. C},
  volume  = {21},
  pages   = {1426--1451},
  year    = {1980}
}

@article{Ji:2000id,
  author        = {Ji, Xiangdong and Lebed, Richard F.},
  title         = {Counting form factors of twist-two operators},
  journal       = {Phys. Rev. D},
  volume        = {63},
  pages         = {076005},
  year          = {2001},
  eprint        = {hep-ph/0012160},
  archiveprefix = {arXiv}
}

@article{Hagler:2004yt,
  author        = {H{\"a}gler, Philipp},
  title         = {Form factor decomposition of generalized parton distributions at leading twist},
  journal       = {Phys. Lett. B},
  volume        = {594},
  pages         = {164--170},
  year          = {2004},
  eprint        = {hep-ph/0404138},
  archiveprefix = {arXiv}
}

@article{Best:1997qp,
  author        = {Best, C. and G{\"o}ckeler, M. and Horsley, R. and Ilgenfritz, E.-M. and Perlt, H. and Rakow, P. and Sch{\"a}fer, A. and Schierholz, G. and Schiller, A. and Schramm, S.},
  title         = {Pion and rho structure functions from lattice {QCD}},
  journal       = {Phys. Rev. D},
  volume        = {56},
  pages         = {2743--2754},
  year          = {1997},
  doi           = {10.1103/PhysRevD.56.2743},
  eprint        = {hep-lat/9703014},
  archiveprefix = {arXiv}
}

@article{Sun:2018bda,
  author        = {Sun, Bao-Dong and Dong, Yu-Bing},
  title         = {Polarized generalized parton distributions and structure functions of the rho meson},
  journal       = {Phys. Rev. D},
  volume        = {99},
  pages         = {016023},
  year          = {2019},
  doi           = {10.1103/PhysRevD.99.016023},
  eprint        = {1811.00666},
  archiveprefix = {arXiv},
  primaryclass  = {hep-ph}
}

@article{Zhang:2022odc,
  author        = {Zhang, Jin-Li and Kang, Guang-Zhen and Ping, Jia-Lun},
  title         = {Rho meson generalized parton distributions in the {Nambu--Jona-Lasinio} model},
  journal       = {Phys. Rev. D},
  volume        = {105},
  pages         = {094015},
  year          = {2022},
  doi           = {10.1103/PhysRevD.105.094015},
  eprint        = {2204.14032},
  archiveprefix = {arXiv},
  primaryclass  = {hep-ph}
}

@book{Varshalovich:1988ye,
  author    = {Varshalovich, D. A. and Moskalev, A. N. and Khersonskii, V. K.},
  title     = {Quantum Theory of Angular Momentum},
  publisher = {World Scientific},
  address   = {Singapore},
  year      = {1988},
  doi       = {10.1142/0270}
}

@article{Burkardt:2000za,
  author        = {Burkardt, Matthias},
  title         = {Impact parameter dependent parton distributions and off-forward parton distributions for $\zeta\to 0$},
  journal       = {Phys. Rev. D},
  volume        = {62},
  pages         = {071503},
  year          = {2000},
  note          = {Erratum: Phys. Rev. D 66, 119903 (2002)},
  doi           = {10.1103/PhysRevD.62.071503},
  eprint        = {hep-ph/0005108},
  archiveprefix = {arXiv}
}

@article{Diehl:2002he,
  author        = {Diehl, Markus},
  title         = {Generalized parton distributions in impact parameter space},
  journal       = {Eur. Phys. J. C},
  volume        = {25},
  pages         = {223--232},
  year          = {2002},
  note          = {Erratum: Eur. Phys. J. C 31, 277--278 (2003)},
  doi           = {10.1007/s10052-002-1016-9},
  eprint        = {hep-ph/0205208},
  archiveprefix = {arXiv}
}

@article{Kim:2024soc,
  author        = {Kim, June-Young and Won, Ho-Yeon and Kim, Hyun-Chul and Weiss, Christian},
  title         = {Spin-orbit correlations in the nucleon in the large-$N_c$ limit},
  journal       = {Phys. Rev. D},
  volume        = {110},
  number        = {5},
  pages         = {054026},
  year          = {2024},
  doi           = {10.1103/PhysRevD.110.054026},
  eprint        = {2403.07186},
  archiveprefix = {arXiv},
  primaryclass  = {hep-ph}
}

\end{document}